\documentclass[usenatbib]{mn2e}
\usepackage{apjfonts,amsfonts,amsmath,amssymb,bm,ctable,verbatim}

\newcommand{\pfh}[1]{{#1}}
\newcommand{\Alf}{{Alfv\'en}}

\newcommand{\bhat}{\hat{{\bf b}}}

\newcommand{\gizmourl}{\href{http://www.tapir.caltech.edu/~phopkins/Site/GIZMO.html}{\url{http://www.tapir.caltech.edu/~phopkins/Site/GIZMO.html}}}
\newcommand{\FIREurl}{\href{http://fire.northwestern.edu}{\url{http://fire.northwestern.edu}}}

\newcommand{\papertwo}{Paper {\small II}}
\newcommand{\paperonetwo}{Papers {\small I} \&\ {\small II}}
\newcommand{\paperthree}{H20}
\newcommand{\etal}{et al.}

\newcommand{\plotside}[2]{\centering \leavevmode \includegraphics[width={#2\textwidth}]{#1}}
\newcommand{\datastatement}[1]{\begin{small}\section*{Data Availability Statement}\end{small}{\noindent #1}\vspace{5pt}}
\newcommand{\acknowledgments}[1]{\begin{small}\section*{Acknowledgments}\end{small}{\noindent #1}\vspace{5pt}}
\makeatletter
\newcommand*{\@rowstyle}{}
\newcommand*{\rowstyle}[1]{\gdef\@rowstyle{#1}\@rowstyle\ignorespaces}
\newcolumntype{=}{>{\gdef\@rowstyle{}}}
\newcolumntype{+}{>{\@rowstyle}}
\makeatother

\title[Galactic Effects of CR Transport]{Effects of Different Cosmic Ray Transport Models on Galaxy Formation}

\author[Hopkins \etal]{
\parbox[t]{\textwidth}{
Philip F.~Hopkins$^1$, 
T.~K.\ Chan$^2$, 
Jonathan Squire$^3$,
Eliot Quataert$^4$, 
Suoqing Ji$^1$, \\
Du\v{s}an Kere\v{s}$^2$,
Claude-Andr{\'e} Faucher-Gigu{\`e}re$^5$
}\vspace*{4pt} \\
$^1$ TAPIR, Mailcode 350-17, California Institute of Technology, Pasadena, CA 91125, USA. E-mail:phopkins@caltech.edu \\
$^2$ Department of Physics, Center for Astrophysics and Space Science, University of California at San Diego, 9500 Gilman Drive, La Jolla, CA 92093 \\
$^3$ Physics Department, University of Otago, 730 Cumberland St., Dunedin 9016, New Zealand \\
$^4$ Department of Astronomy and Theoretical Astrophysics Center, University of California Berkeley, Berkeley, CA 94720 \\ 
$^5$ Department of Physics and Astronomy and CIERA, Northwestern University, 2145 Sheridan Road, Evanston, IL 60208, USA \\ 
}

\date{}
\begin{document}
\maketitle

\begin{abstract}
Cosmic rays (CRs) with $\sim$\,GeV energies can contribute significantly to the energy and pressure budget in the interstellar, circumgalactic, and intergalactic medium (ISM, CGM, IGM). Recent cosmological simulations have begun to explore these effects, but almost all studies have been restricted to simplified models with constant CR diffusivity and/or streaming speeds. Physical models of CR propagation/scattering via extrinsic turbulence and self-excited waves predict transport coefficients which are complicated functions of local plasma properties. In a companion paper, we consider a wide range of observational constraints to identify proposed physically-motivated cosmic-ray propagation scalings which satisfy both detailed Milky Way (MW) and extra-galactic $\gamma$-ray constraints. Here, we compare the effects of these models relative to simpler ``diffusion+streaming'' models on galaxy and CGM properties at dwarf through MW mass scales. The physical models predict large local variations in CR diffusivity, with median diffusivity increasing with galacto-centric radii and decreasing with galaxy mass and redshift. These effects lead to a more rapid dropoff of CR energy density in the CGM (compared to simpler models), in turn producing weaker effects of CRs on galaxy star formation rates (SFRs), CGM absorption profiles and galactic outflows. The predictions of the more physical CR models tend to lie ``in between'' models which ignore CRs entirely and models which treat CRs with constant diffusivity.
\end{abstract}


\section{Introduction}
\label{sec:intro}

\begin{footnotesize}
\ctable[caption={{\normalsize Summary of CR transport models from \paperthree\ (\S~\ref{sec:cr.transport.models}), considered here. \pfh{All models include star formation, stellar feedback, MHD, anisotropic conduction/viscosity, and all (except ``SC:Default'' included for reference) produce $\gamma$-ray signatures consistent with observations across SMC-through-starburst galaxies, and similar MW grammage/residence time/energy density at the solar circle.}}\label{tbl:transport}},center,star
]{=r +l }{ 
}{
\hline\hline
Name & Description  \\
\hline\hline
CD:  & \multicolumn{1}{l}{{\bf Constant-Diffusivity (CD) Models:} constant $\kappa_{\|}=\kappa_{29}\,10^{29}\,{\rm cm^{2}\,s^{-1}}$, 
 with streaming $v_{\rm st,\|}= v_{A} = v_{A}^{\rm ideal} = (|{\bf B}|^{2}/4\pi \rho)^{1/2}$}  \\
\hline
  $\kappa_{29}=3$ & $\kappa_{29}=3$, $v_{\rm st,\|}=v_{A}$: lowest-$\kappa_{\|}$ observationally-allowed constant-$\kappa$ model from \paperonetwo \\
 $\kappa_{29}=30$ & $\kappa_{29}=30$, $v_{\rm st,\|}=v_{A}$: highest-$\kappa_{\|}$ observationally-allowed constant-$\kappa$ model from \paperonetwo \\
\hline\hline
ET:  & \multicolumn{1}{l}{{\bf Extrinsic Turbulence (ET) Models}: $\kappa_{\|}=\mathcal{M}_{A}^{-2}\,c\,\ell_{\rm turb}\,f_{\rm turb}$, with different $f_{\rm turb}$, and $v_{\rm st,\|}=0$  (turbulent scale $\ell_{\rm turb}$, \Alf\ Mach $\mathcal{M}_{A}$)}  \\
 \hline 
 \Alf-Max & $f_{\rm turb} = 1$: \Alf-wave scattering in a GS95 cascade ignoring gyro-averaging and anisotropy terms (otherwise $f_{\rm turb} \gtrsim 1000$) \\ 
 Fast-Max & Fast-mode scattering, neglecting ion-neutral and $\beta>1$ viscous damping (otherwise $f_{\rm turb} \gtrsim 10^{4}$) \\
\hline\hline
SC:  & \multicolumn{1}{l}{{\bf Self-Confinement (SC) Models}: $\kappa_{\|}= c\,r_{\rm L}\,(16/3\pi)\,(\ell_{\rm cr}\,\Gamma_{\rm eff}/v_{A}^{\rm ion})\,(e_{\rm B}/e_{\rm cr})\,f_{\rm QLT}$ from gyro-resonant damping $\Gamma$, $v_{\rm st}=v_{A}^{\rm ion} \equiv v_{A}\,f_{\rm ion}^{-1/2}$}  \\
\hline 
 $f_{\rm QLT}$-100 & $f_{\rm QLT}=100$ with $\Gamma=\Gamma_{\rm in}+\Gamma_{\rm turb}+\Gamma_{\rm LL}+\Gamma_{\rm NLL}$ (ion-neutral, turbulent, linear and non-linear Landau damping all included)   \\ 
\pfh{``Default''} & \pfh{as above with $f_{\rm QLT}=1$ (simplest SC model; observationally disfavored by $\gamma$-ray flux+grammage - listed for reference)} \\ 
\hline\hline
\, & \multicolumn{1}{l}{{\bf Combined Self-Confinement \& Extrinsic-Turbulence (SC+ET) Models}: $\kappa_{\|}^{-1} = \sum \kappa_{i}^{-1}$ (sum scattering rates), $v_{\rm st}=v_{A}^{\rm ion}$} \\ 
\hline 
SC+ET & ``Default'' scattering from \Alf{ic} (ET:\Alf-C00) + fast (ET:Fast-YL04) modes + SC with 100x larger $\Gamma_{\rm turb}$ (SC:$f_{\rm turb}=100$)  \\
\hline
}
\end{footnotesize}

\pfh{It is well-established that CRs represent a non-negligible fraction of the energy and pressure budget in the ISM, and a number of recent theoretical studies incorporating CRs (at dominant $\sim\,$GeV energies) in galaxy simulations have argued that CRs could significantly influence galaxy formation, primarily via their effect enhancing outflows, suppressing inflows, and changing the phase structure of the CGM around galaxies \citep[see e.g.][]{jubelgas:2008.cosmic.ray.outflows,uhlig:2012.cosmic.ray.streaming.winds,Boot13,Hana13,Sale14cos,Simp16,Giri16,Pakm16,wiener:2017.cr.streaming.winds,Rusz17,Jaco18,Buts18,farber:decoupled.crs.in.neutral.gas,holguin:2019.cr.streaming.turb.damping.cr.galactic.winds}.} However, the microphysics of $\sim\,$GeV CR transport, encapsulated in parameters like scattering rates $\nu$, diffusivity $\kappa$, or streaming speeds $v_{\rm st}$ remain deeply uncertain, and previous galaxy simulations have adopted extremely simple parameterizations such as assuming a constant diffusivity and/or streaming at some multiple of the \Alf\ speed. In a series of papers, \citet{chan:2018.cosmicray.fire.gammaray,hopkins:cr.mhd.fire2} (\paperonetwo) and \citet{ji:fire.cr.cgm} used such simplified models to argue that CRs could significantly influence galaxy formation in intermediate through Milky Way (MW)-mass halos at low redshifts ($z\lesssim 1-2$), but that these predictions were {\em most} sensitive to the values of the CR transport parameters adopted (compared to all other CR model variations considered). \pfh{Meanwhile, microphysical models for CR transport motivated by CR scattering from either extrinsic turbulent (ET) fluctuations in magnetic fields, or (in more modern models) self-confinement (SC) via gyro-resonant \Alf-waves self-excited by CR streaming motion predict that these coefficients should be non-linear, complicated functions of a number of plasma properties \citep[see][]{Zwei13,zweibel:cr.feedback.review}, e.g.\ turbulent dissipation rates ($\delta{\bf v}_{\rm turb}^{3}/\ell_{\rm turb}$); CR energy densities ($e_{\rm cr}$) and their gradients; magnetic field strengths ($|{\bf B}|$) or energies ($e_{\rm B}$) or the plasma $\beta$; gas densities ($n$), temperatures ($T$), and ionization fractions ($f_{\rm ion}$); and CR gyro radii ($r_{\rm L}$) or frequencies ($\Omega$).}

\pfh{Recently, \citet{hopkins:cr.transport.constraints.from.galaxies} (\paperthree) presented} the first cosmological galaxy simulations that used CR transport coefficients taken from the more complicated scalings predicted by ET or SC models. \paperthree\ considered dozens of model variants (as well as simpler constant diffusivity or CD models), and compared each in detail to CR observations including $\gamma$-ray measurements from the SMC/LMC/M33/MW/M31 and nearby starburst galaxies, and MW grammage/residence time/rigidity-dependence/ionization rate/CR energy density constraints measured at the solar circle. This showed that only a small subset of these models were observationally allowed. However this subset does differ from CD models in important ways: the transport parameters, being nonlinear functions of local plasma properties, vary locally by orders-of-magnitude on sub-kpc scales in the ISM and CGM, and vary systematically with galaxy mass, redshift, and galacto-centric radius in non-trivial fashion. In this letter, we explore how these differences in CR transport physics can influence  {\em galaxy} properties,  considering a subset of observationally-allowed CD, ET, and SC models from \paperthree.

\begin{figure*}
    \plotside{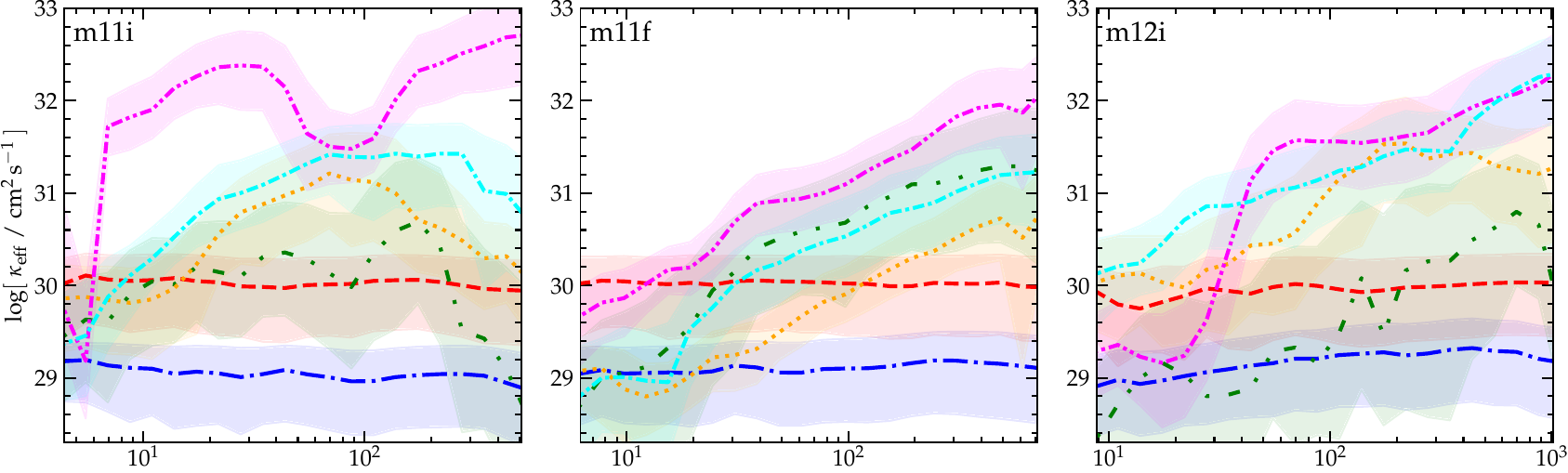}{0.95}\\
    \plotside{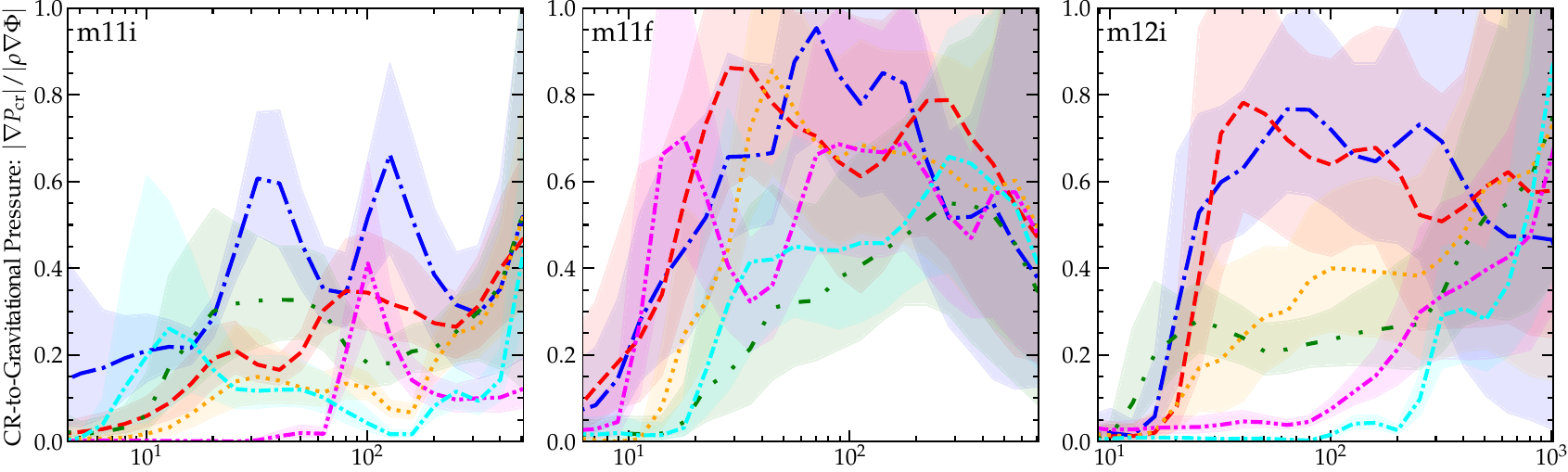}{0.95}\\
    \hspace{-0.15cm}\plotside{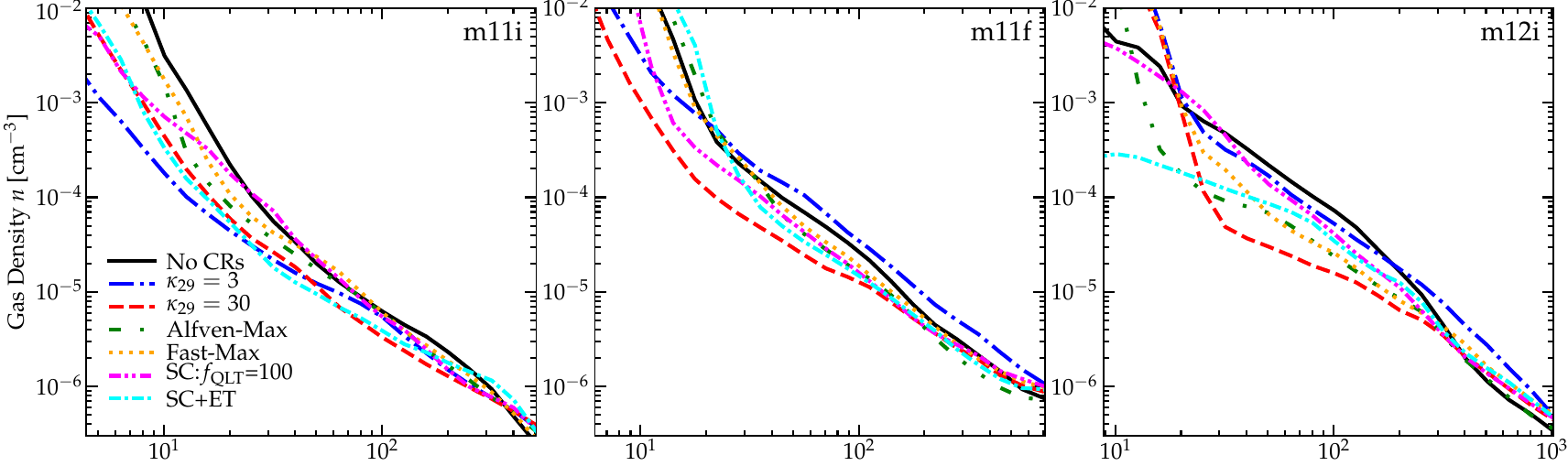}{0.96}\\
    \plotside{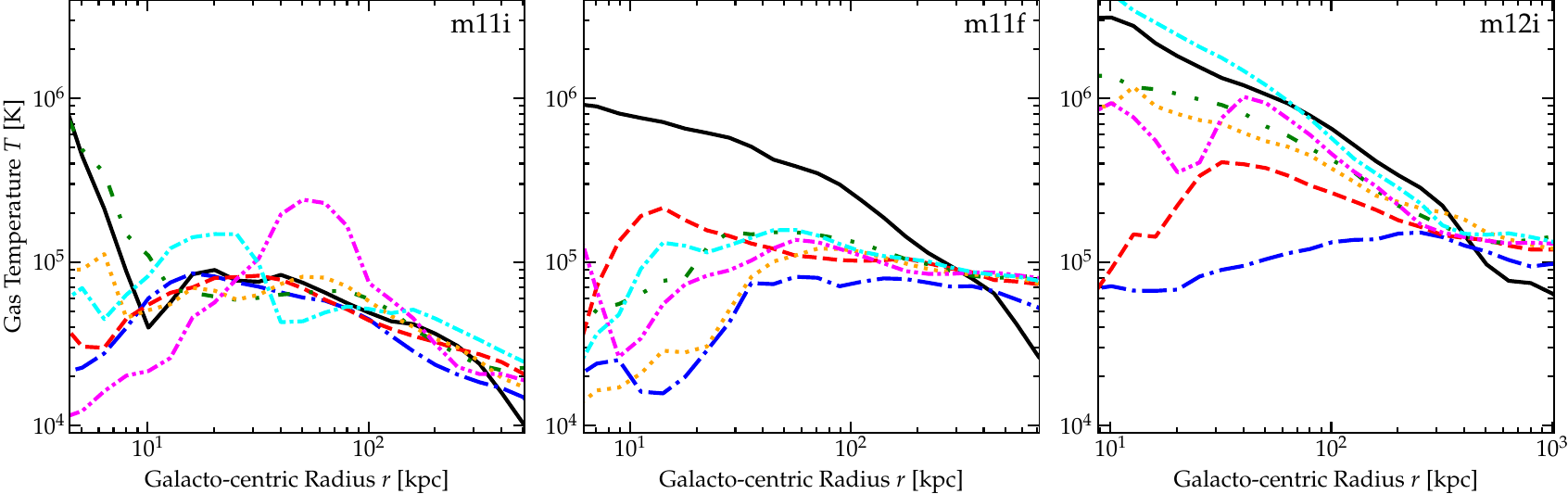}{0.95}\\
    \vspace{-0.2cm}
    \caption{Properties of galaxies {\bf m11i} (SMC-mass dwarf with halo virial mass $M_{\rm vir} \sim 7\times10^{10}\,M_{\odot}$, mass resolution $\Delta m = 7000\,M_{\odot}$), {\bf m11f} (late-type galaxy, intermediate surface-density, $M_{\rm vir}\sim 5\times10^{11}\,M_{\odot}$, $\Delta m = 12000\,M_{\odot}$), {\bf m12i} ($\sim L_{\ast}$ galaxy in massive halo, higher surface-density; $M_{\rm vir} \sim 1.4\times10^{12}\,M_{\odot}$, at lower resolution $\Delta m = 56000\,M_{\odot}$) with different CR transport models (Table~\ref{tbl:transport}), restricted to an ``observationally allowed'' subset (plus ``No CRs,'' for reference), at $z=0$, measured in spherical radial annuli at $r$.
    {\em Top:} ``Effective'' angle-averaged, mean {\em scattering-rate weighted} diffusivity $\kappa_{\rm eff}$ (shaded shows inter-quartile range).
    {\em Second:} Volume-averaged ratio of outward CR pressure force $\nabla P_{\rm cr}$ to gravitational force $\rho\nabla \Phi$. 
    {\em Third:} Volume-averaged gas density $n_{H}$.
    {\em Bottom:} Volume-averaged gas temperature $T$.
    The {\em qualitative} effect of CRs is identical in all cases: the additional CR pressure supports much cooler gas (it does not need to be at the virial temperature), and the additional effect of CR-driven outflows reduces CGM densities (by a small factor). 
    In ET/SC models, $\kappa_{\rm eff}$ varies locally (by $>1$\,dex) but is similar to the allowed CD models {\em within the disk} (where direct CR observations exist) for a subset of the galaxies, but tends to increase with $r$ outside the disk and can increase in other galaxies. This leads to lower CR pressures in the CGM, as (in steady state) $P_{\rm cr} \propto e_{\rm cr} \propto 1/\kappa_{\rm eff}$. This in turn reduces the effects of CRs relative to the CD models.
    \label{fig:cgm.fx}}
\end{figure*}

\section{Methods}
\label{sec:methods}

\subsection{Overview}

The simulations here are all presented in \paperthree, so we briefly summarize here. The simulations are run with {\small GIZMO}\footnote{A public version of {\small GIZMO} is available at \gizmourl} \citep{hopkins:gizmo}, with the meshless finite-mass MFM solver (a mesh-free finite-volume Lagrangian Godunov method). All simulations include ideal magneto-hydrodynamics (MHD; see \citealt{hopkins:mhd.gizmo,hopkins:cg.mhd.gizmo} for methods and tests); anisotropic Spitzer-Braginskii conduction and viscosity (see \citealt{hopkins:cr.mhd.fire2} and \citealt{hopkins:gizmo.diffusion,su:2016.weak.mhd.cond.visc.turbdiff.fx}); and gravity with adaptive Lagrangian force softening for gas (matching the hydrodynamic resolution). These are cosmological ``zoom-in'' runs, evolving a large box from $z\gtrsim 100$ with resolution concentrated on a $\sim1-10\,$Mpc co-moving volume around a ``target'' halo of interest. 

Radiative cooling, star formation and stellar feedback is included following the FIRE-2 implementation of the Feedback In Realistic Environments (FIRE) physics \citep[details in][]{hopkins:fire2.methods}. Cooling from $10-10^{10}$K accounts for metal-line, fine-structure, photo-electric, photo-ionization, cosmic ray, dust, atomic, and molecular processes, including both local radiation sources and the meta-galactic UV background (with self-shielding); we allow star formation only in gas which is locally self-gravitating \citep{hopkins:virial.sf,grudic:sfe.cluster.form.surface.density}, self-shielding, Jeans-unstable, and dense ($>1000\,{\rm cm^{-3}}$). Stars evolve according to standard stellar evolution tracks accounting explicitly for the mass, metals, momentum, and energy injected via individual SNe (Ia \&\ II) and O/B and AGB-star mass-loss \citep[see][]{hopkins:sne.methods} and their radiation (including photo-electric/ionization heating and radiation pressure with a five-band radiation-hydrodynamic scheme; \citealt{hopkins:radiation.methods}). 

The CR treatment follows \paperonetwo, which include extensive tests, as updated in \paperthree. We evolve a single-bin ($\sim$\,GeV) or equivalently constant spectral distribution of CRs as a relativistic fluid\footnote{\pfh{One might question the validity of the fluid approximation as we approach the CR mean free path \citep{2000ifd..book.....B}, and more work on the form and closure of the CR moment equations on these scales is warranted. Note though that CR gyro-radii are much smaller, so our expressions are more akin to kinetic MHD. Moreover as shown in \citet{chan:2018.cosmicray.fire.gammaray,hopkins:cr.mhd.fire2,hopkins:cr.transport.constraints.from.galaxies}, the detailed form of the CR flux equation in this limit has quite weak effects on our conclusions. It is true however that most of our expressions fundamentally assume a gyrotropic CR distribution, which may not always be valid when mean free paths are large.}} (energy density $e_{\rm cr}$, pressure $P_{\rm cr}=e_{\rm cr}/3$; e.g.\ \citealt{mckenzie.voelk:1982.cr.equations}) which obeys the two-moment transport equations:
\begin{align}
\label{eqn:ecr} \frac{\partial e_{\rm cr}}{\partial t}  + \nabla\cdot \left[ {\bf u}\,(e_{\rm cr}+P_{\rm cr}) + {\bf F} \right]
&= {\bf u}\cdot \nabla P_{\rm cr} - \Lambda_{\rm st} - \Lambda_{\rm coll} + S_{\rm in} \\  
\label{eqn:flux} 
\frac{\mathbb{D}_{t}{\bf F}}{\tilde{c}^{2}} + \nabla_{\|} P_{\rm  cr} &= -\frac{{\bf F}}{3\,\kappa_{\ast}} 
\end{align}
where ${\bf u}$ is the gas velocity; ${\bf F}$ the CR flux in the fluid frame; $S_{\rm in}$ the CR source term (determined by assuming $10\%$ of the SNe ejecta energy in each explosion goes into CRs); \pfh{$\nabla_{\|} P_{\rm cr} \equiv \bhat\,(\bhat\cdot \nabla P_{\rm cr})$ is the cosmic-ray pressure gradient parallel to the magnetic field ${\bf B}$;\footnote{\pfh{We neglect perpendicular diffusion, as it is expected to be smaller by orders of magnitude (powers of the ratio of the gyro radius to CR mean free scattering length; see e.g.\ \citealt{rogrigues:cr.diffusion.parallel.perp}).}} $\Lambda_{\rm st}={\rm min}(v_{A},\,v_{\rm st})\,|\nabla_{\|}P_{\rm cr}|$ represents ``streaming losses'' as gyro-resonant \Alf\ waves (with \Alf\ speed $v_{A}$ and unresolved wavelengths of order the CR gyro-radius $r_{\rm L}$) are excited by CR streaming and rapidly damp \citep{wentzel:1968.mhd.wave.cr.coupling,kulsrud.1969:streaming.instability};} $\Lambda_{\rm coll}=5.8\times10^{-16}\,{\rm s^{-1}\,cm^{3}}\,(n_{\rm n}+0.28\,n_{e})\,e_{\rm cr}$ represents collisional losses (with $n_{\rm n}$ the nucleon number densities for hadronic losses, and $n_{e}$ the free electron number density for Coulomb losses; \citealt{guo.oh:cosmic.rays}); $\tilde{c}$ is the maximum (physical or numerical) CR free-streaming/signal speed; $\mathbb{D}_{t}{\bf F} \equiv \hat{\bf F}\,[\partial |{\bf F}|/\partial t + \nabla\cdot({\bf u}\,|{\bf F}|) +  {\bf F}\cdot\{ (\hat{\bf  F}\cdot \nabla)\,{\bf u}\} ]$ is  the derivative operator derived (for a gyrotropic CR distribution function) \pfh{in \citet{thomas.pfrommer.18:alfven.reg.cr.transport};} and $\kappa_{\ast}\equiv \kappa_{\|} + (4/3)\,\ell_{\rm cr}\,v_{\rm st,\,\|}$ (with $\ell_{\rm cr}\equiv P_{\rm cr}/|\nabla_{\|} P_{\rm cr}|$ the CR gradient scale-length) defines the streaming/diffusive speeds. CRs influence the gas as the appropriate fraction of the streaming+hadronic+Coulomb losses are thermalized while the CR pressure appears in the gas momentum equation.

\begin{figure*}
    \plotside{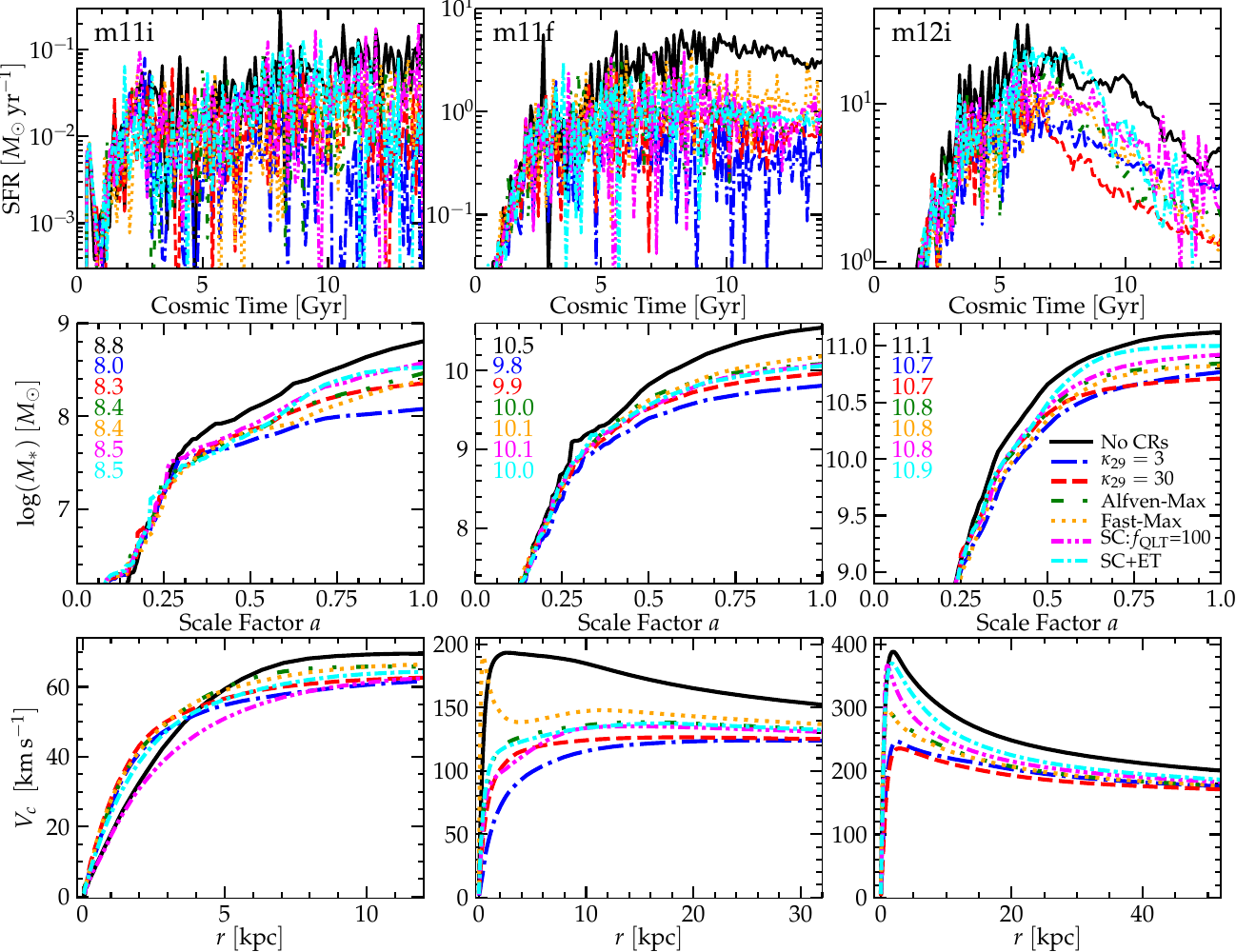}{0.99}
    \vspace{-0.25cm}
    \caption{Galaxy properties for {\bf m11i}, {\bf m11f}, {\bf m12i}. 
    {\em Top:} Star formation (SF) history (averaged in $100\,$Myr bins) of the galaxy, vs.\ cosmic time.
    {\em Middle:} Stellar mass inside the virial radius vs.\ scale-factor ($a=1/(1+z)$), with $z=0$ value shown (number).
    {\em Bottom:} \pfh{Circular velocity $V_{c}^{2}\approx G\,M_{\rm enc}(<r)/r$ vs.\ (spherical) galactocentric radius $r$ at $z=0$.}
    In all cases, CRs suppress SF at $z\lesssim 1-2$ in these galaxies. 
    Constant-diffusivity models tend to over-estimate this suppression relative to ET or SC models, 
    as (per \papertwo) CR feedback is primarily ``preventive'' via suppressing CGM inflows.
    \label{fig:gal.fx}}
\end{figure*}

\subsection{Different CR Transport Models}
\label{sec:cr.transport.models}

With these definitions, $\kappa_{\|}$ and $v_{\rm st,\,\|}$ are the traditional CR parallel diffusivity and streaming speeds: a ``pure diffusion'' equation for CR transport results if we take $\tilde{c}\rightarrow \infty$ (eliminating the second-moment/flux equation) with $\kappa_{\|}=\,$constant and $v_{\rm st,\,\|}=0$, while a ``pure streaming/advection'' equation results from $\tilde{c}\rightarrow\infty$ with $\kappa_{\|}=0$ and $v_{\rm st,\,\|}=\,$constant. In \paperonetwo, we show that the overwhelmingly dominant uncertainty in modeling  CRs and their effects on galaxies is the choice of $\kappa_{\ast}$ (i.e.\ $\kappa_{\|}$ and $v_{\rm st,\,\|}$). Other parameters in our CR model are uncertain at a more modest (factor $\sim 2$) level, and produce only weak effects on galaxy properties: in contrast $\kappa_{\ast}$ varies by up to $\sim 8$ orders of magnitude in different models discussed in the literature for the MW alone (see \paperthree) and qualitatively changes whether or not CR pressure can couple to gas at all. Here we consider only a small subset of models (summarized in Table~\ref{tbl:transport}) from \paperthree\ that can reproduce the CR observations. The full scalings of $\kappa_{\|}$ and $v_{\rm st}$ with plasma properties are quite complicated and are presented in full in \paperthree, we simply summarize.

\begin{enumerate}

\item {\bf Constant diffusivity (CD) models} adopt $\kappa_{\|}=\,$constant and $v_{\rm st}=v_{A}^{\rm ideal}$ (ideal MHD \Alf\ speed). These have no particular physical motivation, but are the most commonly adopted models in the literature and were the basis of \paperonetwo\ (where we showed CR observations require $\kappa_{\|} \sim (3-30)\times10^{29}\,{\rm cm^{2}\,s^{-1}}$). 
\item {\bf Extrinsic Turbulence (ET) models} assume CR scattering from pre-existing turbulence.  The most theoretically well-motivated models from \paperthree\ (``\Alf-C00'' and ``Fast-YL04'')  produce far too-low a scattering rate (too-high $\kappa$) compared to $\sim$\,GeV CR observations, which likely indicates that  self-confinement dominates $\sim\,$GeV CR scattering. To explore the somewhat uncertain turbulent scattering physics, \paperthree\ did consider variant ET models that involved either ignoring gyro-averaging effects from anisotropy\footnote{\pfh{Per \citet{chandran00}, the ``anisotropy'' referred to here in the context of models like ``\Alf-Max'' is the anisotropy of the \Alf{ic} modes in the extrinsic turbulent cascade at small (gyro-resonant) scales, which is predicted to be substantial in any scenario which exhibits critical balance (e.g.\ a \citealt{GS95.turbulence}-type cascade). The CRs themselves assumed to be nearly isotropic.}} (model ``\Alf-Max''), or ignoring some fast-mode damping terms (model ``Fast-Max''). These, although poorly motivated theoretically, provide a useful reference because the dimensional scaling of $\kappa$ resembles that of the most well motivated models. This scaling, which, for ``\Alf-Max,'' is $\kappa_{\|}\sim \mathcal{M}_A^{-2} c\, \ell_{\rm turb}\, f_{\rm turb}$ (with \Alf\ Mach number $\mathcal{M}_A$ on  turbulent length scale $\ell_{\rm turb}$), shows that $\kappa$ should be expected to increase at large radii due to the increase in turbulent length scales. Other effects, which are absorbed in $f_{\rm turb}$ and included in other turbulence models (e.g., $f_{\rm turb}\sim30\,\mathcal{M}_A^{5/3}\,{\rm Re}^{-1/3}$ in the simplest regime for ``Fast-Max,'' where ${\rm Re}$ is the Reynolds number), do not change this expectation.

\item {\bf Self-confinement (SC) models}  assume CRs are scattered due to self-excited gyro-resonant \Alf-waves, and so depend on gas and CR parameters differently from ET models. \paperthree\ showed that \pfh{the simplest}, theoretically well-motivated SC models \pfh{(e.g.\ the model ``SC:Default'' in Table~\ref{tbl:transport})} tend to over predict scattering rates (under-predict $\kappa$) by a factor $\sim100$. \pfh{The discrepancy with theoretical expectations could be accounted for by a number of significant uncertainties in SC physics (see \S~5.3 of \paperthree); we therefore consider both this and the models from \paperthree\ which adopt the quasi-linear ``correction factor'' $f_{\rm QLT}=100$ to match observations. }The dependence of $\kappa_{\|}$ on plasma properties is complicated in these models (see Table~\ref{tbl:transport}), and depends on ionization fraction $f_{\rm ion}$, $e_{\rm cr}$ and $\ell_{\rm cr} \equiv e_{\rm cr}/|\nabla_{\|} e_{\rm cr}|$, and damping rates $\Gamma$ of gyro-resonant modes from a variety of sources (see \paperthree\ for details): if e.g.\ turbulence dominates damping ($\Gamma\approx \Gamma_{\rm turb}$), then $\kappa_{\|} \propto \ell_{\rm cr}\,e_{\rm cr}^{-1}\,|\delta{\bf v}_{\rm turb}|^{3/2}\,\rho^{3/4}\,\ell_{\rm turb}^{-1/2}\,f_{\rm ion}^{1/2}\,f_{\rm QLT}$. In general because $\kappa_{\|} \propto \ell_{\rm cr}/e_{\rm cr}$ in SC, the falloff of $e_{\rm cr}$ (and increase in $\ell_{\rm cr}$) \pfh{with (spherical) galacto-centric radius} generally produce rising $\kappa_{\|}$. 

\item {\bf Combined ``SC+ET'' models} assume scattering rates from SC and ET models, adopting the \Alf-C00 and Fast-YL04 models above and (for the sake of contrast) a slightly-different SC model with the ``quasi-linear'' correction applied to the turbulent damping rate $\Gamma_{\rm turb}$ as compared to the total $\Gamma$ or $\kappa$ (as in $f_{\rm QLT}$-100). 

\end{enumerate}

\section{Results}

\subsection{Effects on the CGM}

Fig.~\ref{fig:cgm.fx} illustrates the key differences between the allowed models in the CGM. We compare three halos (a representative dwarf, intermediate and massive system from \paperthree), at $z=0$, with the CR models from Table~\ref{tbl:transport}, and a reference ``No CRs'' model which does not include CRs. We measure the angle-averaged (isotropic-equivalent) mean CR scattering-rate-weighted diffusivity $\kappa_{\rm eff} \equiv \int \,(|{\bf F} \cdot \hat{r} | / | \nabla e_{\rm cr} |)\,d\varpi$ (with weight $d\varpi = (e_{\rm cr}/\kappa_{\|})\,d^{3}{\bf x} / \int (e_{\rm cr}/\kappa_{\|})\,d^{3}{\bf x}$), the ratio of the CGM radial pressure gradient to the inward gravitational force, and CGM gas density and temperature in radial annuli. 

\paperthree\ studied the diffusivities $\kappa_{\rm eff}$ and transport parameters in detail -- showing for example that the ET/SC models plotted here produce order-of-magnitude local variations in $\kappa_{\|}$ that in turn mean the diffusivity weighted by scattering rate or volume or other parameters can be quite distinct. Here, what is notable is how the diffusivities tend to increase with $r$ outside the galaxy. Recall, all models here are constrained to reproduce the same CR observations ($\gamma$-ray luminosities, grammage, residence/decay times, etc.) -- but these observables are entirely dominated by gas within the central $\lesssim 10\,$kpc around the galaxy. As a result, the allowed models tend to reproduce similar scattering-weighted mean $\kappa$ in the ISM (at least for some MW-mass systems) but extrapolate differently. The SC/ET models show $\kappa$ rising with $r$, as $n$, $|{\bf B}|$, $\delta {\bf v}_{\rm turb}^{3}/\ell_{\rm turb}$, and other relevant quantities which drive CR scattering decrease. \pfh{Interestingly, because the SC models feature a more sharply-rising $\kappa_{\rm eff}$ outside the galaxy, the model ``SC:Default'' from \paperthree\ (with $f_{\rm QLT}=1$, so $\kappa_{\rm eff}$ is systematically lower by a factor $\sim 100$ compared to ``SC:$f_{\rm QLT}=100$''), while it gives very low $\kappa_{\rm eff}$ {\em within} the galaxy (in potential tension with observations), gives quite similar $\kappa_{\rm eff}$ in the CGM to the ET models considered here.}

In \papertwo, we show that in steady-state with some fixed CR injection rate in the galaxy $\dot{E}_{\rm cr}$, neglecting losses (as the $\gamma$-ray observations require weak losses for the galaxies here), the CR pressure in the CGM must scale as $P_{\rm cr} \sim \dot{E}_{\rm cr} / (12\pi\,\kappa_{\rm eff}\,r)$. Thus, the higher-$\kappa$ outside the galaxy in SC/ET models leads to lower $P_{\rm cr}$, which in turn means a lower gas density $\rho \sim |\nabla P_{\rm cr}| / | \nabla \Phi |$ can be supported by CR pressure against gravity (set primarily by the dark matter). As a result we see either (a) gas fails to be supported by CRs ($|\nabla P_{\rm cr}| / | \rho\,\nabla \Phi | \ll 1$) or (b) lower CGM gas densities (smaller $n$) appear when $\kappa_{\rm eff}$ is larger outside the galaxy. \papertwo\ also shows that effects of CRs in the CGM become weaker in low-mass halos, as CR pressure (given much lower star formation rates) is less able to support the halo gas (lower $|\nabla P_{\rm cr}| / | \rho\,\nabla \Phi | $), producing overall weaker effects of CRs in {\bf m11i}.

\citet{ji:fire.cr.cgm} further explored the CGM properties of the $\kappa_{29}=3$ runs here in detail. They showed that for runs {\bf m11f} and {\bf m12i}, where CRs in this run dominate the pressure support against gravity, gas at relatively low temperature $T$ can be supported, while in the ``No CRs'' run, the dominant pressure support is thermal, requiring the gas have temperature of order the virial temperature (or else it falls onto the galaxy and is accreted rapidly). This leads directly to different CGM temperatures in Fig.~\ref{fig:cgm.fx}, which in turn translates to different metal ionization states (with the cooler CGM in the $\kappa_{29}=3$ run primarily photo-ionized, and the hotter CGM in the ``No CRs'' run collisionally ionized) and therefore different equivalent widths in common CGM metal absorption lines. We see that with higher CGM diffusivities, the SC/ET models generally produce results ``in between'' the CD and ``No CRs'' models here. \pfh{For this reason noted above, SC models from \paperthree\ with lower $f_{\rm QLT}\sim1$, for this reason, end up close to the ``\Alf-Max'' or ``Fast-Max'' models in the CGM properties in Fig.~\ref{fig:cgm.fx}.}

\subsection{\pfh{Effects on Galaxy Evolution \&\ the ISM}}

Fig.~\ref{fig:gal.fx} examines the effects of different CR transport models on galaxy evolution (comparing to a baseline simulation with no CRs). This is the main focus of \paperonetwo, which explore a much wider variety of galaxy properties, for constant-diffusivity models. However \papertwo\ shows that almost all of the effects of CRs are strongly-correlated because the effect of CRs is primarily as a ``preventive'' feedback agent, contributing to CGM pressure and slowly-accelerating outflows, preventing cool gas from more-rapidly accreting onto the galaxy. Thus to the extent that CRs suppress the SFRs/stellar masses in galaxies, they also modify the galaxy morphologies, metallicities, angular momentum content, ISM phases, rotation curves, etc, but simply ``moving along'' the correlations of these quantities with galaxy stellar mass and/or SFR. We therefore restrict in Fig.~\ref{fig:gal.fx} to cosmological SFRs, the buildup of stellar mass, and galaxy rotation curves (concentration, creation of ``cores,'' etc).

\papertwo\ showed that the range $\kappa_{29}\sim 3-30$ favored by $L_{\gamma}$ observations is also the range of $\kappa$ which produces the {\em maximal} effect of CRs on intermediate-through-MW mass systems here. At much lower-$\kappa$, CRs lose all their energy collisionally, so provide no pressure and do essentially nothing to the galaxy. At much higher-$\kappa$, CRs simply ``escape'' and produce negligible pressure/coupling to gas. At the ``sweet spot'' in $\kappa$, CRs from SNe can reduce SFRs by as much as a factor $\sim 5$ near $z\sim0$, and stellar masses by factors $\sim 2-4$, acting primarily at low redshifts/late times (as shown in \papertwo). However, we find that in the more physically-motivated SC and ET models, the effects of CRs on galaxy evolution are weaker. They are still appreciable, but almost always lie ``between'' the ``No CRs'' and ``Constant Diffusivity'' models. \pfh{For the reasons above, with its intermediate diffusivity in the CGM, the reference model ``SC:Default'' with $f_{\rm QLT}\sim1$ produces results similar here to the intermediate models, closes to ``Fast-Max'' in {\bf m11i} and {\bf m11f} and to ``SC+ET'' in {\bf m12i}.} \pfh{This follows naturally from the arguments above: if the diffusivity rises with galacto-centric radius in these models, the CR pressure drops more rapidly, so CRs are less efficient at re-accelerating winds and suppressing accretion onto galaxies, compared to constant-diffusivity models.}

\pfh{As noted, we do not explicitly consider ISM properties here, though they will be studied in much greater detail comparing the effects of some of the CR models here in Chan et al. (in prep) and Benincasa et al. (in prep). To leading-order, the effects of CRs shift galaxies ``along'' a sequence via their effects in the CGM, changing the baryonic mass of the galaxy. But also recall that these models are all calibrated to reproduce a similar set of observational constraints including $\gamma$-ray luminosities and grammage and CR energy density at the solar circle in MW-like galaxies. As a result, properties like the diffusivity and CR pressure are, by construction, quite similar in the disk midplane. While the models predict quite distinct variations across different phases, since they have similar ``mean'' behavior, these generally produce second-order differences.}

\section{Conclusions}
\label{sec:conclusions}

We consider for the first time the effect on galaxy and CGM properties of different physically-motivated CR transport models including diffusion and streaming coefficients that vary with local plasma properties (motivated by micro-physical CR transport models), in fully-cosmological, multi-phase ISM/CGM galaxy formation simulations. All the models here are constrained to reproduce similar CR observables including $\gamma$-ray luminosities of dwarf, MW-like, and starburst galaxies, MW grammage and residence time and CR energy density/ionization rate constraints (\paperthree). However these observations  only significantly constrain CR transport coefficients within the galactic ISM and innermost CGM (radii $\lesssim 10$\,kpc). 

In the physically-motivated models, there are large local variations in the ``effective'' CR diffusivity $\kappa_{\rm eff}$, and $\kappa_{\rm eff}$ tends to increase significantly in the CGM, because CR scattering rates decrease in the lower-density, higher-$\beta$ gas. This increasingly rapid CR diffusion leads to a more rapidly decreasing CR pressure as a function of galacto-centric radius $r$, which limits the range of CGM/IGM radii over which CRs can contribute significantly to supporting gas against gravity (relative to models with a constant CR diffusivity). That, in turn, means CRs produce weaker effects on CGM phase structure, density and temperature profiles, outflow re-acceleration, and suppression of galactic star formation. As a result, in all galaxy and CGM properties examined here or in  \paperonetwo, and \citet{ji:fire.cr.cgm}, the {\em qualitative} effects of CRs are similar, but the physically motivated CR transport models tend to produce {\em quantitative} effects in-between our runs without CRs and our favored ``constant diffusivity'' runs. This suggests the effects of CRs on galaxy formation, while not negligible, are relatively modest, altering galaxy masses by factors up to $\sim 2-3$, and temperatures in the CGM around $\sim L_{\ast}$, low-redshift galaxies by factors up to $\sim 2-5$. 

The major caveat of this study is the uncertainty regarding the true microphysical CR transport model. We showed in \paperthree\ that the models here are observationally allowed, but not unique. Moreover, our conclusion in \paperthree\ was that neither  ET nor  SC models can reproduce all relevant CR observations with their simplest  ``default'' parameterizations, but that they require some modifications in order to fit the observations. For example, the SC model we consider here scales the scattering rate by the  arbitrary factor $f_{\rm QLT} \sim 100$, which is necessary in order to avoid over-confining CRs \pfh{{\em in the ISM}}. However, these modifications (required to match observations) are not necessarily well-understood theoretically, and it is possible that they could also vary systematically with radius. If the ``correct'' $f_{\rm QLT}$ were not constant but instead $\sim 100$ inside the galactic disk \pfh{(or even within just certain parameter-space regimes relevant for $\gamma$-ray production in our simulations)} and decreased with $\propto 1/r$ outside the disk, then the SC model here would still reproduce  observations, \pfh{but with a diffusivity much more weakly dependent on galacto-centric radius $r$, thus more closely resembling the ET or CD models.} Our hope is that the combination of CGM, galaxy, and direct CR observations may better constrain these parameters in the future, and (in the meantime) these different results allow us to better understand the dominant systematic uncertainties in theoretical models that attempt to predict the effects of CRs on galaxy formation.

\datastatement{The data supporting the plots within this article are available on reasonable request to the corresponding author. A public version of the GIZMO code is available at \gizmourl. Additional data including simulation snapshots, initial conditions, and derived data products are available at \FIREurl.}

\acknowledgments{\pfh{We thank the anonymous referee for a number of insightful comments.} Support for PFH was provided by NSF Collaborative Research Grants 1715847 \&\ 1911233, NSF CAREER grant 1455342, and NASA grants 80NSSC18K0562 and JPL 1589742. CAFG was supported by NSF 1517491, 1715216, and CAREER 1652522; NASA 17-ATP17-0067; and by a Cottrell Scholar Award. Support for JS  was provided by Rutherford Discovery Fellowship RDF-U001804 and Marsden Fund grant UOO1727 from the Royal Society Te Ap\=arangi. DK was supported by NSF grant AST-1715101 and the Cottrell Scholar Award from the Research Corporation for Science Advancement. Numerical calculations were run on the Caltech compute cluster ``Wheeler,'' allocations FTA-Hopkins supported by the NSF and TACC, and NASA HEC SMD-16-7592.}

\bibliography{ms_extracted}

\end{document}